\documentclass[aps,prl,twocolumn]{revtex4}
\usepackage[dvips]{graphics}
\usepackage[dvips]{graphicx}
\usepackage{amsmath,amsthm,amsfonts,amssymb} 

\begin{document}
\title{Schr\"odinger-Newton equation as a possible generator of quantum state reduction}
\author{Jasper van Wezel$^1$ and Jeroen van den Brink$^{2,3}$}
\affiliation{
$^1$Cavendish Laboratory, University of Cambridge, Madingley Road, Cambridge CB3 0HE, UK \\
$^2$Institute-Lorentz for Theoretical Physics, Universiteit Leiden,
P.O. Box 9506, 2300 RA Leiden, The Netherlands\\
$^3$Institute for Molecules and Materials, Radboud Universiteit Nijmegen,
P.O. Box 9010, 6500 GL Nijmegen, The Netherlands
}
\date{\today}

\begin{abstract}
It has been suggested by Di\'osi and Penrose that the occurrence of quantum state reduction in macroscopic objects is related to a manifestation of gravitational effects in quantum mechanics. Although within Penrose's framework the dynamics of the quantum state reduction is not prescribed, it was suggested that the so called Schr\"odinger-Newton equation can be used to at least identify the resulting classical end states. Here we analyze the extent to which the Schr\"odinger-Newton equation can be used as a model to generate a full, time dependent description of the quantum state reduction process. We find that when supplied with an imaginary gravitational potential, the Schr\"odinger-Newton equation offers a rationalisation for some of the hitherto unexplained characteristics of quantum state reduction. The description remains incomplete however, because it is unclear how to fully recover Born's rule.
\end{abstract}
\maketitle

\section{I: Introduction} \noindent
More than a decade ago Penrose and, independently, Di\'osi noticed that the inherent incompatibility of quantum mechanics' unitary time evolution with general relativity's diffeomorphism invariance could have measurable consequences on mesoscopic scales~\cite{Penrose96,diosi87}. Although the precise way in which general relativity would alter quantum mechanics on these scales is not known theoretically, and proves to be extremely hard to access experimentally~\cite{Marshall03,Christian05,vanWezel08}, Penrose and Di\'osi use very general arguments to suggest that the process might be used to clarify the mystery of quantum state reduction. In a nutshell, the main argument is that gravity necessarily induces some uncertainty in the total energy of a quantum superposition state and thus allows quantum superpositions only a finite lifetime given by the inverse of that energy uncertainty. In turn, this can in principle lead to testable predictions about the timescale at which quantum state reduction is expected to occur~\cite{Penrose96,diosi87}. On the other hand, this argument does not provide information on the dynamics of the reduction process itself. The equations that describe the transition from a well defined microscopic superposition state to a collapsed macroscopic measuring apparatus remain unknown. Progress in this direction was made by proposing to use the so called Schr\"{o}\-dinger-Newton equation as the defining equation for which states could be stable under gravitationally induced collapse, and which could not~\cite{Moroz98,diosi84}. This Schr\"{o}\-dinger-Newton equation is a non-linear set of equations defined as
\begin{eqnarray}
-\frac{\hslash^2}{2m} \nabla^2 \psi + U \psi = E \psi   \nonumber \\
\nabla^2 U = 4 \pi G m^2 \left| \psi \right|^2,
\label{SNeq}
\end{eqnarray}
where $\psi$ is the wavefunction for a particle with mass $m$, $\hbar$ Planck's constant, $U$ the gravitational potential, $E$ the energy eigenvalue and $G$ the gravitational constant. Because the total energy operator is the generator for time translations, we should in principle also be able to use these equations to describe the dynamical reduction of a quantum superposition. Here we investigate the possibility of taking the Schr\"{o}dinger-Newton equation~\eqref{SNeq} not only as an equation governing the final shape of gravitationally collapsed states, but also as a generalization of the quantum mechanical Hamiltonian operator, which can give us both the total energy and the time evolution of a quantum superposition state.

In our study of the possible implied time evolutions, we will adopt the philosophy that in order for the Schr\"{o}dinger-Newton equation to be considered as a candidate description for the generator of quantum state reduction, it should at least be able to correctly describe the simplest possible situations in which quantum state reduction is expected to occur. We will thus calculate its implied dynamics in a number of simplified settings, without considering whether these situations correspond to practical experiments. The rationale is that if the dynamics does not yield the desired outcome in these simple cases, we should not expect it to do so in more complex real-life situations.

In the next section we will start out by examinig the dynamics of a system defined in a Hilbert space that consists of only two possible states. We will see that the Schr\"{o}dinger-Newton equation in that case does not gives rise to quantum state reduction, but that it can be easily adjusted into a related equation which does give the desired result. Next, in section III, we turn to the description of a superpostion over three initial states. Here even the adjusted equation is seen to harbor only two of the three necessary characteristics for a full description of reduction dynamics. In section IV we will thus conclude that although the Schr\"{o}dinger-Newton equation does offer a rationalisation for some of the hitherto unexplained characteristics of quantum state reduction, it can nonetheless not be the full story.

\section{II: A Two State Measurement} \noindent
The very simplest possible experiment that any description of a quantum state reduction process should be able to describe is the evolution of a macroscopic superposition over two distinct states to just one of these states. One can imagine the macroscopic superposition to be formed in a process in which we start out with some microscopic superposition of two different quantum states. The difference between these states might give rise to a gravitational self-energy, but the mass involved is supposed to be so small that the microscopic matter will not collapse by itself. At a certain point in time a coupling between the microscopic state and a macroscopic measuring machine will be instantaneously turned on. In general this will yield a macroscopic superposition and the difference between the superposed states will generically have a finite gravitational self-energy. Now the macroscopic mass involved in this self-energy is expected to make the collapse process very fast, and as a result it will seem as if an instantaneous measurement has yielded only one of the two possible outcomes. According to Born's rule~\cite{Born26} the distribution between the two outcomes found in many repetitions of the experiment must mirror the squared wavefunction of the microscopic superposition that we started out with.

\subsection{The General Two State Time Evolution}
To see whether the Schr\"{o}dinger-Newton equation can indeed lead to the collapse of a two state measurement, we will first write down the generic time evolution of a superposition of two states. The most general superposition state over a basis with two elements is given by
\begin{eqnarray}
\left| \psi_0 \right> = n e^{i \frac{\chi}{2}} \left[ e^{i \frac{\varphi}{2}} \cos \left( {\theta/2} \right) \left|0 \right> + e^{-i \frac{\varphi}{2}} \sin \left( {\theta/2} \right) \left|1 \right> \right],
\end{eqnarray}
where $n$ is the norm of the wavefunction, which is usually set to $1$, and where $\chi$ is the total phase which is usually ignored because it cannot be measured by any quantum mechanical process. To define the time evolution of this wavefunction we introduce a generator for time translations $G$~\cite{Sakurai}. In its most general form, this generator is a complex $2$x$2$ matrix in the basis $\{ \left|0\right>, \left|1\right> \}$, with eight independent real entries:
\begin{eqnarray}
G = \left( \begin{array}{cc} \alpha_R + i \alpha_I & \beta_R + i \beta_I \\ \gamma_R + i \gamma_I & \delta_R + i \delta_I \end{array} \right).
\end{eqnarray}
The Hermitian part of this generator will coincide with the usual Hermitian quantum mechanical Hamiltonian $H$. The remaining non-Hermitian (and possibly even non-linear) part is used to generate the non-unitary collapse dynamics. The time evolution of the wavefunction can be generated from the definition of $G$ by looking at the infinitesimal time translation of $\left| \psi_0 \right>$ \index{collapse}
\begin{align}
\left| \psi_{\epsilon} \right> &= e^{i \epsilon G} \left| \psi_0\right> \nonumber \\
&= \left( 1 + i \epsilon G + O \left( \epsilon^2 \right) \right) \left| \psi_0 \right> \nonumber \\
\equiv N &e^{i \frac{X}{2}} \left[ e^{i \frac{\varPhi}{2}} \cos \left( {\Theta/2} \right) \left|0 \right> + e^{-i \frac{\varPhi}{2}} \sin \left( {\Theta/2} \right) \left|0 \right> \right].
\label{time}
\end{align}
Here $\epsilon$ is an infinitesimally small parameter that measures time in units of $\hslash$. The parameters $N$, $X$, $\varPhi$ and $\Theta$ defining the new wavefunction at time $t= \epsilon$ are written in terms of the old parameters $n$, $\chi$, $\varphi$ and $\theta$ at time $t=0$, and are defined to first order in $\epsilon$. It is now easy to extract the time evolution of these parameters using the definition of the time derivative. The time derivative of for example $n$ is given by the limit $\epsilon \to 0$ of $\left(N-n\right)/\epsilon$. After some algebra this implies that
\begin{align}
\dot{\theta} =& \left(\alpha_I - \delta_I\right) \sin \left( \theta\right) \nonumber \\
& +2 \left( \beta_I \cos \left( \varphi \right) - \beta_R \sin \left( \varphi \right) \right) \sin^2\left(\theta / 2\right) \nonumber \\
& -2 \left( \gamma_I \cos \left( \varphi \right) + \gamma_R \sin \left( \varphi \right) \right) \cos^2\left(\theta / 2\right) \nonumber \\
\dot{\varphi} =& \left(\alpha_R - \delta_R\right) + \left( \beta_R \cos \left( \varphi \right) + \beta_I \sin \left( \varphi \right) \right) \tan \left(\theta / 2\right) \nonumber \\
& - \left( \gamma_R \cos \left( \varphi \right) - \gamma_I \sin \left( \varphi \right) \right) \cot \left(\theta / 2\right) \nonumber \\
\dot{\chi} =& \left(\alpha_R + \delta_R\right) + \left( \beta_R \cos \left( \varphi \right) + \beta_I \sin \left( \varphi \right) \right) \tan \left(\theta / 2\right) \nonumber \\
& + \left( \gamma_R \cos \left( \varphi \right) - \gamma_I \sin \left( \varphi \right) \right) \cot \left(\theta / 2\right) \nonumber \\
\dot{n} =& -\alpha_I \cos^2\left(\theta / 2\right) - \delta_I \sin^2\left(\theta / 2\right) \nonumber \\
& + {1/2}\left( \beta_R-\gamma_R\right) \sin\left(\varphi\right) \sin\left( \theta \right) \nonumber \\
& - {1/2}\left( \beta_I+\gamma_I\right) \cos\left(\varphi\right) \sin\left( \theta \right).
\label{flow}
\end{align}
Notice that in the case of purely unitary time evolution, generated by a purely Hermitian generator $G$, the derivatives simplify considerably, and become identical to the usual quantum mechanical time evolution
\begin{eqnarray}
\dot{\theta} &=& 2 \left( \beta_I \cos \left( \varphi \right) - \beta_R \sin \left( \varphi \right) \right) \nonumber \\
\dot{\varphi} &=& \left(\alpha_R - \delta_R\right) -2 \left( \beta_R \cos \left( \varphi \right) + \beta_I \sin \left( \varphi \right) \right) / \tan \left(\theta\right) \nonumber \\
\dot{\chi} &=& \left(\alpha_R + \delta_R\right) +2 \left( \beta_R \cos \left( \varphi \right) + \beta_I \sin \left( \varphi \right) \right) / \sin \left(\theta\right) \nonumber \\
\dot{n} &=& 0.
\end{eqnarray}

\subsection{Specific Time Evolutions}
\begin{figure}[tb]
\includegraphics[width=0.3\columnwidth]{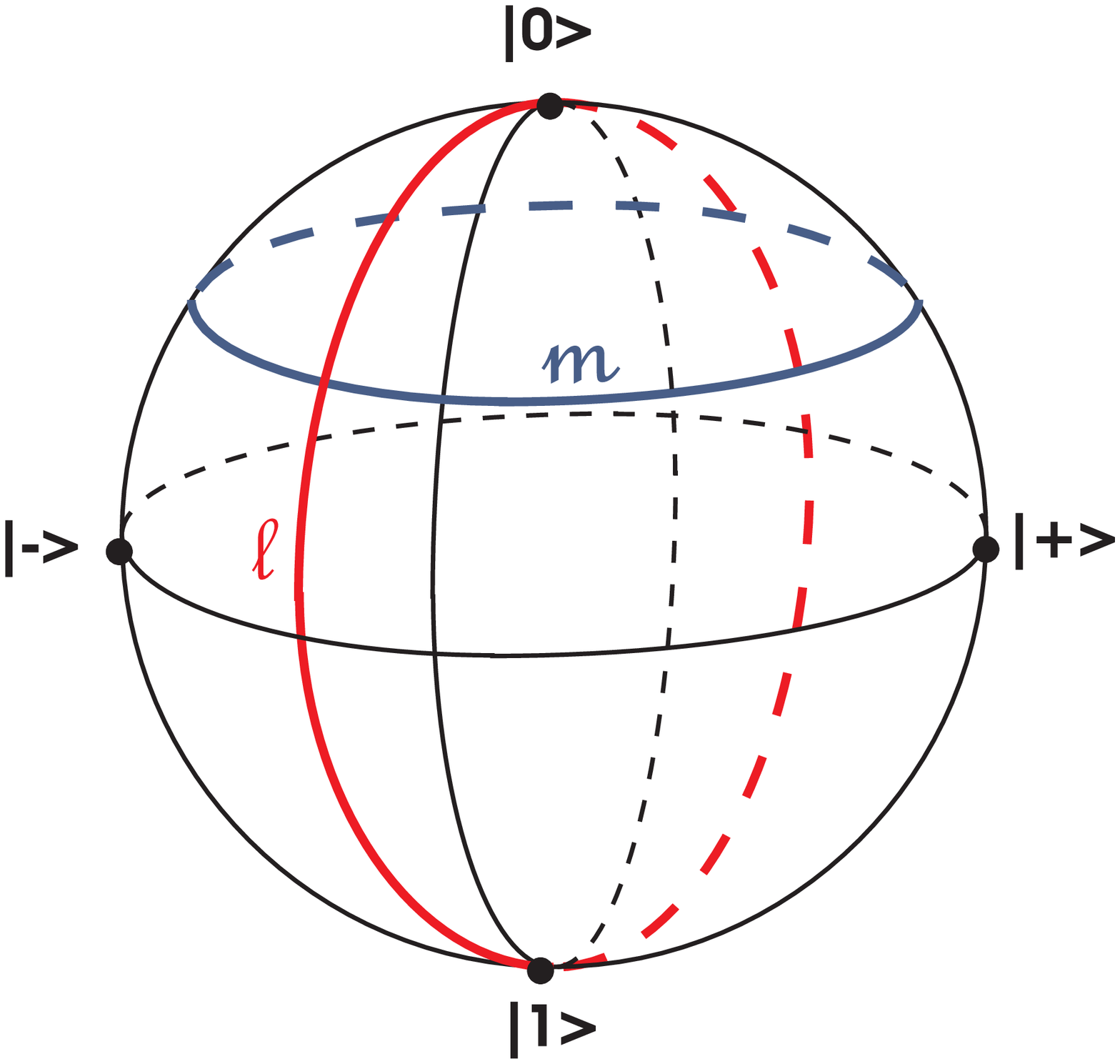}
\hspace{5pt}
\raisebox{7pt}{\includegraphics[width=0.6\columnwidth]{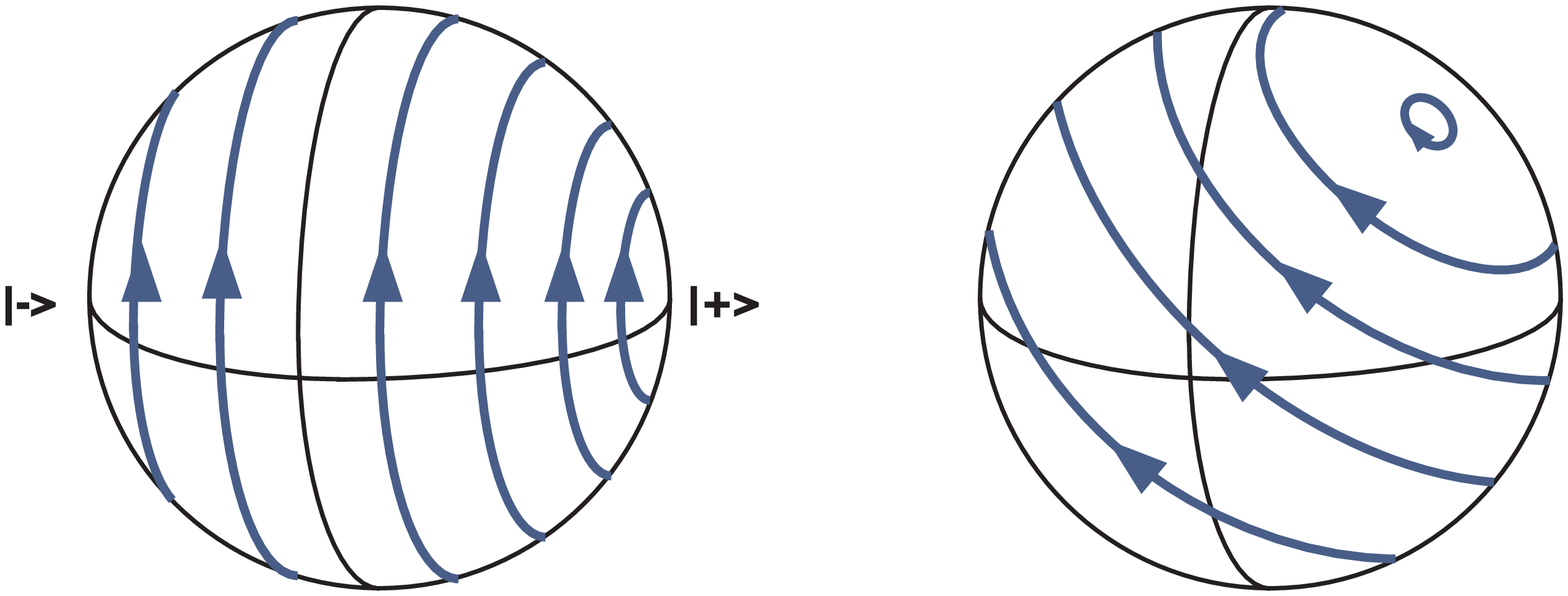}}
\caption{Left: the Bloch sphere. The line $m$ is a line of constant $\theta$, while $l$ has constant $\varphi$. The states $\left| \pm \right>$ are defined as $\sqrt{ 1/2} \left[ \left|0\right> \pm \left|1\right> \right]$. Middle: the flow as described by equations~\eqref{flow1}. Right: the flow pattern of some general Hamiltonian generator of time evolution.}
\label{Bloch}
\end{figure}

It is clear from the time derivatives~\eqref{flow} that the total phase and norm variables do not influence the time evolution of the superposition state as long as the generator $G$ does not explicitly depend on them. We can thus study the time evolution by considering only the variables $\varphi$ and $\theta$. Their time evolution can be visualized as a flow on the Bloch sphere (see figure~\ref{Bloch}). Each flowline on the sphere then represents the path traced out by the time evolution of an initial state somewhere along the path. As long as the time evolution is purely unitary, and the generator $G$ thus purely Hermitian, the flow pattern is in fact always the same: it consists of rotations around an axis spanned by the two eigenstates of the Hamiltonian, as depicted in figure~\ref{Bloch}. These eigenstates are always on opposite poles, and their rotation away from the north and south pole depends on the rotation of the Hamiltonian away from being a diagonal operator.

\subsubsection{Hamiltonian Flow}
As an example, let us consider the Hamiltonian with $\beta_R = \gamma_R = \Delta$ and all other entries zero. 
%
\begin{eqnarray}
\dot{\theta} &=& -2  \Delta \sin \left(\varphi\right) \nonumber \\
\dot{\varphi} &=& - 2  \Delta \cos \left(\varphi\right) / \tan \left(\theta\right) \nonumber \\
\dot{\chi} &=& 2  \Delta \cos \left(\varphi\right) / \sin \left(\theta\right) \nonumber \\
\dot{n} &=& 0.
\label{flow1}
\end{eqnarray}
As mentioned before, we are really only interested in the explicit time evolution of $\varphi$ and $\theta$. To find a closed form for the description of the flowlines it is useful to notice that $\partial_{\theta} \left[ \dot{\theta} \sin\left( \theta\right) \right] = - \partial_{\varphi} \left[ \dot{\varphi} \sin\left( \theta\right) \right]$. This implies that the set of differential equations that we are trying to solve is a so called exact set of ordinary differential equations~\cite{Ross}, and that we can solve it by looking for a potential $V$ which obeys
\begin{eqnarray}
-\partial_{\theta} V &=& \dot{\varphi} \sin\left( \theta\right) \nonumber \\
\partial_{\varphi} V &=& \dot{\theta} \sin\left( \theta\right).
\end{eqnarray}
This set of equations is easily solved, and yields the potential $V=2 \Delta \cos(\varphi)\sin(\theta)$. The streamlines describing the flow on the Bloch sphere\index{Bloch sphere} are lines of constant potential $V$, which are given by
\begin{eqnarray}
\cos \left(\varphi\right) \sin \left(\theta\right) = \text{ constant}.
\end{eqnarray}
The flow therefore is a rotation around the axis through the north and south poles at $\sqrt{ 1/2} \left[ \left|0\right> \pm \left|1\right> \right]$, as seen in figure~\ref{Bloch}. These are of course also precisely the eigenstates of the Hamiltonian. The flow in circles around the poles (for example the one starting out at $\left|0\right>$) correspods to what one usually refers to as Rabbi oscillations.
%
%

\subsubsection{Schr\"{o}dinger-Newton}
\begin{figure}[tb]
\center
\includegraphics[width=0.73\columnwidth]{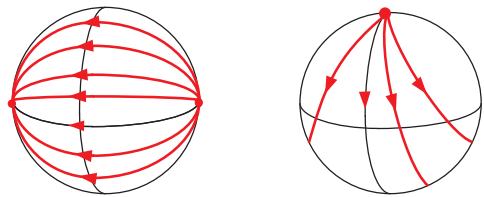}
\caption{Left: the flow of the type $\alpha_I=-\delta_I \neq 0$. Right: the flow of the type $\beta_R \neq 0, ~\gamma_R =0$.}
\label{nonHam}
\end{figure}

In addition to the Hamiltonian part of $G$ we could also include non-Hermitian terms in  the definition of the generator of time evolution. In general these terms can lead to many different possible flow patterns on the Bloch sphere. To get an idea of the types of patterns that can be constructed this way, let us consider three of the most basic possibilities for creating purely non-unitary dynamics.
If we set all entries in the $2$x$2$ matrix $G$ equal to zero except for the terms $\alpha_I$ and $\delta_I$, the resulting time evolution does not alter the individual components of the superposed state, and only influences the norm of the wavefunction. On the other hand, evolutions of the type generated by $\alpha_I=-\delta_I \neq 0$ and rotations thereof (like $\beta_R=-\gamma_R$ or $\beta_I=\gamma_I$) cause a flow from one pole to the opposite pole. That is, one of the poles becomes a source and the opposite a sink for the flowlines. Finally, asymmetric terms like $\beta_R \neq 0, ~\gamma_R =0$, give rise to asymmetric flows through a saddle point with flowlines both emerging from them and disappearing into them, as shown in figure~\ref{nonHam}.

The specific non-unitary pattern we are interested in here, is of course the one given by the non-linear generator implied by the Schr\"{o}dinger-Newton equation~\eqref{SNeq}~\cite{Moroz98,diosi84}. The time evolution generator introduced in the Schr\"{o}dinger-Newton equation for the specific case of a superposition of mass over two distinct positions $\left|x=0\right>$ and $\left|x=1\right>$ is
\begin{eqnarray}
G = \left( \begin{array}{cc} U\left(x=0\right) &  \Delta \\  \Delta & U\left(x=1\right) \end{array} \right),
\label{G1}
\end{eqnarray}
where the gravitational potential $U(x)$ is given by $\nabla^2 U(x) = \gamma \left| \left< \psi \right. \left| x \right> \right|^2$. For simplicity we will absorb the constant $\gamma$ into the norm of the wavefunction, and again ignore the dynamics of that norm. We will also take out the normal Hamiltonian part of $G$ by setting $ \Delta=0$. After all, the reduction process should be caused by the non-unitary part of the time evolution.
\begin{figure}[tb]
\center
\includegraphics[width=0.9\columnwidth]{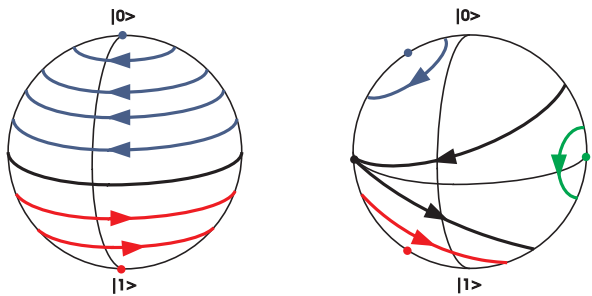}
\caption{Left: the flow as defined by the generator $G$ of equation~\eqref{G1}, with $ \Delta$ set to zero. Right: the flow defined by the Schr\"{o}dinger-Newton equation, including the kinetic energy.}
\label{SNflow}
\end{figure}

To solve for the gravitational potential while avoiding the infinite self-energy of a point particle, we will consider the states labeled by $x$ to represent a mass distribution stretched out over an infinite sheet in the $y,z$-plane, but completely localized in the $x$-direction. The superposition thus achieved should be a good description of for example a large block of mass which is superposed over a distance small compared to its own length. Using appropriate boundary conditions for the gravitational potential and completely ignoring the norm of the wavefunction, the essential part of the time evolution generator reduces to
\begin{eqnarray}
G = \left( \begin{array}{cc} -\cos \left( \theta \right) & 0 \\ 0 & \cos \left( \theta \right) \end{array} \right).
\end{eqnarray}
Clearly the matrix $G$ is a non-linear operator, because it depends on the value of the parameter $\theta$ which defines the state on which $G$ acts. On the other hand $G$ is still akin to a Hermitian matrix in the sense that its transpose equals its complex conjugate. The flow pattern associated with this generator is easily found to consist of circular flowlines around the north and south pole. In contrast with the usual Hamiltonian flow though, the circulation on the northern hemisphere is in the opposite direction of its southern counterpart (see figure~\ref{SNflow}). The gravitational term has thus introduced a division between the northern and the southern hemisphere, but it has not caused any sinks or sources to appear on the Bloch sphere, and it is thus inadequate as a description of the dynamical quantum reduction process.
Even if we reintroduce the mixing parameter $ \Delta$, this will only distort the flow lines from their perfectly circular orbits and produce some sort of a tennis ball flow pattern as depicted in figure~\ref{SNflow}. However, it does not introduce any sources or sinks that would represent the final states in a collapse process.

\subsubsection{Alternative Gravitational Terms}
The lack of sources and, more importantly, sinks in the flow pattern associated with the Schr\"{o}dinger-Newton equation implies that it cannot be used as a description of the dynamical process of quantum state reduction. We can however try and introduce sinks into the dynamics by slightly altering the Schr\"{o}dinger-Newton equation. The simplest way to do so, as was also suggested by Di\'osi~\cite{diosi07}, is to turn the gravitational self-energy term into an imaginary potential, so that the equation becomes
\begin{figure}[tb]
\center
\includegraphics[width=0.9\columnwidth]{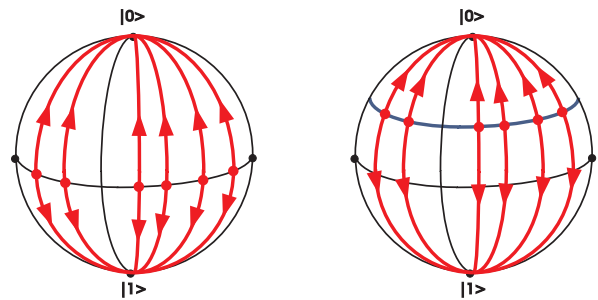}
\caption{Left: the flow as defined by the generator $G'$ of equation~\eqref{G2}. Right: the flow defined by the adjusted generator $G''$ of equation~\eqref{G3}.}
\label{collFlow}
\end{figure}
\begin{eqnarray}
-\frac{\hslash^2}{2m} \nabla^2 \psi + i U \psi = E \psi   \nonumber \\
\nabla^2 U = 4 \pi G m^2 \left| \psi \right|^2.
\label{SNeq2}
\end{eqnarray}
The exact physical meaning of the imaginary potential term is for the moment unclear, but its mathematical form suggests the possibility of referring to it as a dissipation term induced into quantum mechanics by gravity. For now we will not consider the justification of the non-unitary dynamics, but merely study the implications of having a generator of time evolution given by
\begin{eqnarray}
G' = \left( \begin{array}{cc} - i \cos \left( \theta \right) & 0 \\ 0 & i \cos \left( \theta \right) \end{array} \right).
\label{G2}
\end{eqnarray}
The flowlines generated by this matrix all lie along the meridian and the flow goes north on the northern hemisphere, while it goes south on the southern hemisphere (see figure~\ref{collFlow}). Thus all states starting out above the equator will eventually collapse onto the north pole, and all states south of the equator find their destination on the south pole. If we introduce a normal, real quantum mechanical energy into the dynamics as well, then these straight flowlines turn into spirals which flow around the north-south axis as well as toward one of the poles. 

The dynamics has now two of the three properties expected of a working model for quantum state reduction. It identifies the states into which a superposition can collapse (the poles) by making sure that spatial superpositions disappear. At the same time it explains why microscopic superpositions can exist while macroscopic superpositions are never seen: the spiraling motion for microscopic particles is extremely close to perfect circular motion because the mass of the particles is small compared to their usual quantum mechanical potential and kinetic energy. On the other hand the gravitational term dominates for macroscopic objects, and thus their superposition states will be destroyed in a very short time.
In its present form, however, the time evolution defined by $G'$ cannot be used to reproduce the third requirement: Born's rule. A state on the northern hemisphere will always collapse onto the north pole and never onto the south pole. The only way to cure this problem is to introduce a random variable into the dynamics~\cite{diosi07}. In a description of the state reduction process on the level of the Schr\"{o}dinger-Newton equation the origin of this random variable remains unclear. It could correspond to an elusive, as yet unnoticed, field which only manifests itself on the edge between quantum mechanics and gravity. An alternative justification could be to identify the random variable with the total phase of the wavefunction. We then effectively use the total phase of a quantum mechanical state  as a random variable. This seems natural as it is an unmeasurable quantity in quantum mechanics. After introducing it into the altered Schr\"{o}dinger-Newton equation, the time evolution generator becomes
\begin{eqnarray}
G'' = \left( \begin{array}{cc} - i \left[ \cos \left( \theta \right) + f\left(\chi\right) \right] & 0 \\ 0 & i \left[ \cos \left( \theta \right)+ f\left(\chi\right) \right] \end{array} \right).
\label{G3}
\end{eqnarray}
Here $f\left( \chi \right)$ is a function of the random variable $\chi$ that will be fixed by the requirement that the outcome of the dynamics agree with Born's rule. The flow pattern of this adjusted $G''$ is the same as the flow pattern we found before for $G'$, only now the seperatrix between streaming northward and streaming southward lies not at the equator but at the line $\cos\left(\theta\right)=-f\left(\chi\right)$ (see figure~\ref{collFlow}). If we assume $\chi$ to be taken at random from a flat distribution between $0$ and $2 \pi$, then it is easily checked that the dynamics agrees perfectly with Born's rule \index{Born's rule}if we set $f\left(\chi\right)=\chi/\pi-1$.

We have thus found a model that $(i)$ describes the collapse of a quantum mechanical superposition over two different states, $(ii)$ distinguishes between microscopic and macroscopic superpositions and $(iii)$ results in the emergence of Born's rule if it is repeated many times with the same initial conditions, apart from a single random variable that remains unobservable.

\section{III: A Three State Measurement} \noindent
Although the model seems to work well for describing the reduction dynamics of a two-state measurement, it cannot be related to any possible scenario for the solution of the measurement problem unless it also works for more general superpositions. The first step toward testing the model for such a general situation is to ensure that it works for a wavefunction superposed over three states instead of just two. To do so we repeat the analysis of the previous section using the initial state
\begin{align}
\left| \psi_0 \right> = n e^{i \frac{\chi}{2}} & \left[ e^{i \frac{\varphi+\phi}{2}} \cos \left( {\theta/2} \right) \cos \left( {\eta/2} \right) \left|0 \right> \right. \nonumber \\
 & + \left. e^{i \frac{\varphi-\phi}{2}} \cos \left( {\theta/2} \right) \sin \left( {\eta/2} \right) \left|1 \right> \right.  \nonumber \\
 & + \left. e^{-i \frac{\phi-\varphi}{2}} \sin \left( {\theta/2} \right) \left|2 \right> \right]. 
\end{align}
\begin{figure}[tb]
\center
\includegraphics[width=0.9\columnwidth]{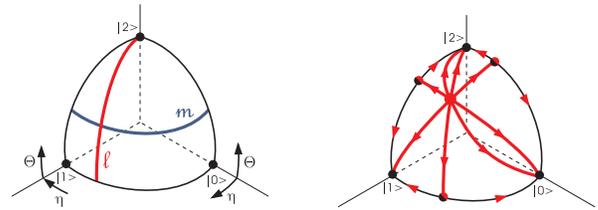}
\caption{Left: the quarter of a sphere on which the three-state time evolution can be depicted. The line $m$ is a line of constant $\theta$ while $l$ has constant $\eta$. Right: the generic flow pattern for the flow defined by equations~\eqref{3flow}.}
\label{quart}
\end{figure}
If we also use a $3$x$3$ matrix for the time evolution generator, then the computation of the time derivatives of $\theta$, $\eta$ and so on is exactly analogous to the two state case. To see what the effect of our modified Schr\"{o}dinger-Newton time evolution $G''$ is in this case, we consider the states $\left|0\right>$, $\left|1\right>$ and $\left|2\right>$ to represent infinitely high columns of mass, positioned on the vertices of an equilateral triangle in the $x,y$-plane. That way we find that the gravitational potential, up to constant prefactor, is given by
\begin{eqnarray}
\nabla^2 U \left( \left| x\right> \right) &\propto& \left|\left< \psi \right|\left. x\right>\right|^2 \nonumber \\
\Rightarrow U\left( \left| 0\right> \right) &\propto& 1 - 3 \cos^2 \left( {\theta/2} \right) \cos^2 \left( {\eta/2} \right) \nonumber \\
U\left( \left| 1\right> \right) &\propto& 1 - 3 \cos^2 \left( {\theta/2} \right) \sin^2 \left( {\eta/2} \right) \nonumber \\
U\left( \left| 2\right> \right) &\propto& 1 - 3 \sin^2 \left( {\theta/2} \right).
\end{eqnarray}
Using these values we can then define the generator of time evolution $G''$ in analogy with equation~\eqref{G3} by setting $\left<x\right| G'' \left| x \right> = i \left( U  \left( \left| x\right> \right) + f_x \left( \chi \right) \right)$. In analogy to the two state time evolution, we have introduced a function $f$ which depends on a single random variable and which is designed to generate the randomness we need to be able to agree with Born's rule. The time derivatives which describe the flow of the state vector through configuration space during the time evolution defined by this $G''$ are then
\begin{align}
\dot{\theta} = \sin \left( \theta \right) & \left[ a \cos^2 \left( {\eta/2} \right) + b - \cos \left( \theta \right) + \right. \nonumber \\
& \frac{1}{2}  \left. \cos^2 \left( {\theta/2} \right) \sin^2 \left( \eta \right) \right] \nonumber \\
\dot{\eta} = \sin \left( \eta \right) &\left[ a - \cos^2 \left( {\theta/2} \right) \cos \left( \eta \right) \right],
\label{3flow}
\end{align}
with $a\equiv f_0 -f_1$ and $b\equiv f_1 - f_2$. 
The relative phases $\varphi$ and $\phi$ are constant in time as long as we do not consider any off-diagonal elements in $G''$, and are therefore irrelevant for the reduction process. To visualize the flow we can use the surface of one octant of a sphere on which $\theta$ measures altitude and $\eta$ latitude, so that the states $\left|0\right>$, $\left|1\right>$ and $\left|2\right>$ are at the vertices of the surface, as shown in figure~\ref{quart}. The generic flow diagram of the equations~\eqref{3flow} has a central source from which all flowlines emanate. The flowlines end either at the sinks located at the vertices of the octant or at a saddle point on one of the edges of the surface. From the saddle point the flow continues to the vertices again (see figure~\ref{quart}).

Changing the values of $a$ and $b$ corresponds to moving the position of the central source over the entire surface and at the same time shifting the saddle points along the edges. To be precise, the position of the fixed points are given in terms of $\left(\eta, \theta\right)$ coordinates as
\begin{align}
P_C &= \left( \arccos\left(\frac{3a}{a+2b+2}\right) , \arccos\left(\frac{1+2a+4b}{3}\right) \right) \nonumber \\
P_{0-1} &= \left( \arccos \left( a \right) , 0 \right) \nonumber \\
P_{1-2} &= \left( \pi, \arccos \left( b \right) \right) \nonumber \\
P_{2-0} &= \left( 0, \arccos \left( a + b \right) \right).
\end{align}
Here $P_C$ is the central source and $P_{i-j}$ is the saddle point on the edge connecting $\left|i\right>$ with $\left|j\right>$. Clearly the flow pattern for the three state superposition fulfills the first two requirements for being considered as a description of quantum state reduction. The stable points (the sinks) in the flow represent precisely the three possible wavefunctions that do not involve a superposition over gravitationally distinct states, and that are therefore acceptable as possible outcomes of a quantum measurement. The time involved in getting to such a stable state is again governed by the ratio between gravitational and kinetic energy. The microscopic superpositions will thus be able to avoid collapse for a very long time, while macroscopic superpositions are doomed to collapse within moments after their creation.

\subsection{Born's Rule}
The challange that is left for the three-state problem is to find a function $f$ such that repeated application of the measurement model~\eqref{3flow} yields Born's rule. As before the introduction of some random variable cannot be avoided, and again we will try and use just one random variable, which could be either an unobserved field or the total phase of the wavefunction.

\begin{figure}[tb]
\center
\includegraphics[width=.9\columnwidth]{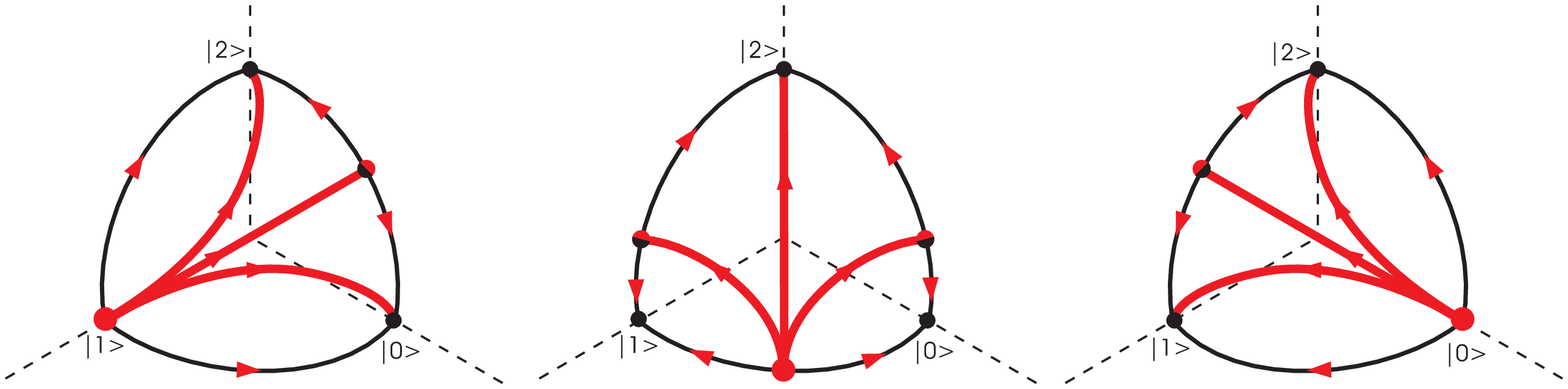}
\caption{A few of the flow diagrams that are encountered as $\chi$ moves from $0$ to $2\pi$ in equations~\eqref{f}.}
\label{sideFlow}
\end{figure}

When proposing an Ansatz for $f$ we should keep in mind that the three-state time evolution must reduce to the two-state time evolution which we found before in the case that the initial state happens to be on one of the edges of configuration space. This in fact implies that the saddle points on the edges must move along the edges for varying $\chi$ precisely like the seperatrix moved along the meridian on the two-state Bloch sphere. In addition we ought to demand that the collection of all possible flow patterns posses a 3-fold rotational symmetry in the sense that for every flow pattern in the collection there must be two more flow patterns which coincide with the original one if the vertices are interchanged in a cyclic fashion. In the end there seems to be just one possible choice for the function $f$ (or equivalently, for $a$ and $b$) that satisfies all of these conditions and depends only on one random variable $\chi$. This somewhat pathlogical looking choice is given by
\begin{eqnarray}
a &=& \left\{ \begin{array}{cc} -1 + \frac{3 \chi}{2 \pi} & \text{ if } ~\chi < \frac{4 \pi}{3} \\ 5-\frac{3\chi}{\pi} & \text{ if } ~\chi \geq \frac{4 \pi}{3} \end{array} \right. \nonumber \\
b &=& \left\{ \begin{array}{cc} 1 - \frac{3 \chi}{\pi} & \text{ if } ~\chi < \frac{2 \pi}{3} \\ -2 +\frac{3\chi}{2 \pi} & \text{ if } ~\chi \geq \frac{2 \pi}{3} \end{array} \right. .
\label{f}
\end{eqnarray}
These forms for $a$ and $b$ imply that as a function of $\chi$ the central source moves all around the perimeter of configuration space, while the saddle points move up and down their respective edges, as depicted in figure~\ref{sideFlow}.
This ensures that Born's rule will hold on the edges of configuration space, as it did in the two-state superposition scenario. Whether or not it holds away from the edges is difficult to prove analytically because an equation for the flow lines connecting the central source to the saddle points is not easily found. Numerically however it is rather straightforward to just simulate the reduction dynamics many times and compare the result with the expected outcome based on Born's rule.

\begin{figure*}[tb]
\center
\includegraphics[width=0.95\columnwidth]{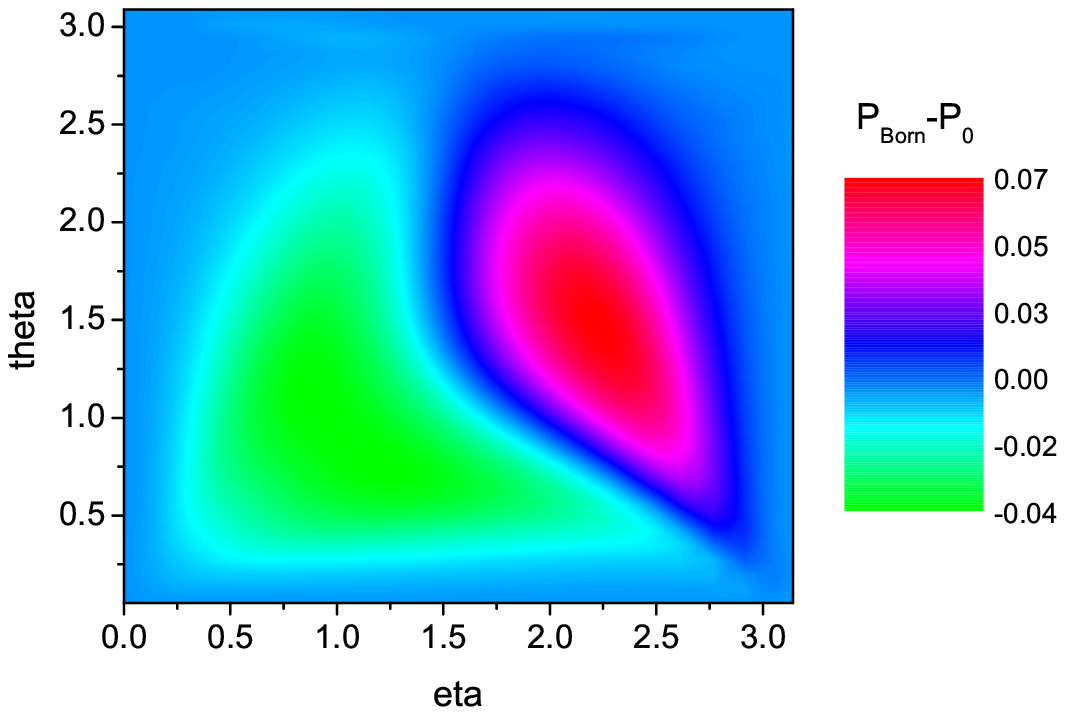}
\includegraphics[width=0.95\columnwidth]{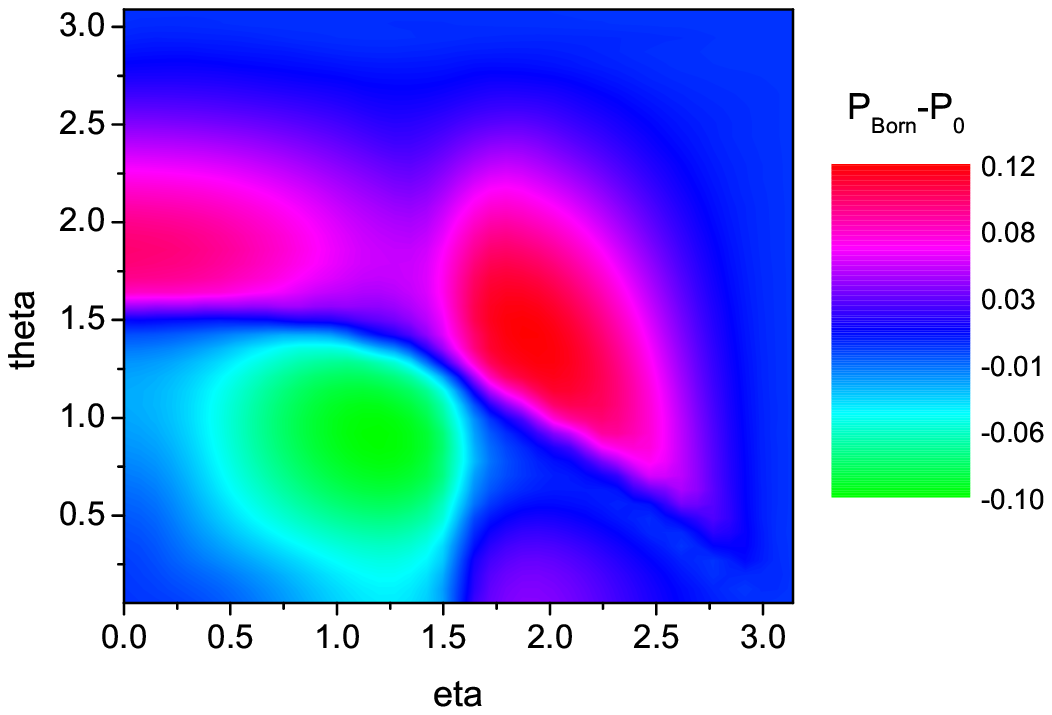}
\caption{False color plot of the difference between the probability for finding outcome $\left|0\right>$, following the time evolution $G''$ from the initial state ($\eta$,$\theta$) and the corresponding probability according to Born's rule (i.e. $\cos^2(\theta/2) \cos^2(\eta/2)$). \\
Left: Using the measurement scheme defined by equations~\eqref{f}, in which the source of the flow wanders around all edges. \\ 
Right: Using the measurement scheme in which we force the centre of the flow to be in a random position for each value of $\chi$.}
\label{bornplotSW}
\end{figure*}
%
%
%

As it turns out, the proposed dynamics, including the definitions~\eqref{f}, do not agree with Born's rule. The difference is shown in figure~\ref{bornplotSW}. To fix the mismatch one could try other ways to define $f$. We could look for a different scheme in which $f$ satisfies all necessary conditions but differs from~\eqref{f}; we could introduce a dependence of $f$ on $\eta$ or $\theta$; or we could introduce additional random variables. We  did not find any of these approaches to be viable. Even the simulated solution in which we force the position of the central source to form a flat distribution over configuration space in the course of many experiments does not yield the desired result (see figure~\ref{bornplotSW}). Moreover, the introduction of more random variables would be a rather undesirable element in the theory, because they should physically emerge from ever more fluctuating fields for whose existence there is no empirical evidence.

\section{IV: Conclusions} \noindent
The results discussed in this paper show that the Schr\"{o}dinger-Newton equation~\eqref{SNeq} which was proposed by Penrose as a replacement for the quantum mechanical Hamiltonian~\cite{Moroz98,diosi84} can hardly serve as a description of the dynamics of quantum state reduction. On the other hand, a slight modification of the equation, i.e. making the gravitational potential energy appear as an imaginary term, causes the associated dynamics to show at least two of the three characteristics necessary for being considered as a possible reduction model. The modified equation causes the system to evolve toward states which are not superposed over gravitationally different positions, and thus selects the correct Pointer basis for the quantum system to collapse into\cite{diosi07,Zurek81}. On top of this the equation naturally provides a reason for the observed difference between microscopic and macroscopic objects. Microscopic systems have a very small gravitational potential energy as compared to their internal quantum mechanical potential and kinetic energy. The collapse process will therefore be so slow that it cannot be noticed on human timescales. On the other hand the gravitational term will dominate in the dynamics of macroscopic superposition states, and these will thus collapse before their existence can be noticed~\cite{Penrose96,diosi87}.

If the wavefunction is a sum of only two distinct states then the addition of a random variable into the dynamics rather straightforwardly leads to the desired statistics for the outcomes of measurements. However, as soon as the wavefunction represents a superposition over more then two states, it becomes impossible to force the dynamics of the collapse model to agree with Born's rule using only one random variable. Even apart from the fact that there is no physical ground for introducing them, more than one random variable does not automatically solve the problem. At least the obvious choices of how to implement them into the theory do not seem to yield the desired outcome.

We therefore conclude that the Schr\"{o}dinger-Newton equation, even in its modified forms, is still not fit as a complete description of the dynamical collapse process of quantum superpositions, because it is unclear how to make it agree with Born's rule under any but the most basic circumstances.

\subsection{Acknowledgements} \noindent
We thank Jan Zaanen for numerous discussions and gratefully acknowledge support from the Dutch Science Foundation FOM.


\end{document}